\documentclass{article}
\usepackage{icmc2020template}
\usepackage{times}
\usepackage{ifpdf}
\usepackage{soul}
\usepackage[english]{babel}

\usepackage{booktabs}
\usepackage{multirow}
\usepackage{amsmath}
\usepackage{tabularx}

\def\papertitle{Modeling Baroque Two-Part Counterpoint with Neural Machine Translation}
\def\firstauthor{Eric P. Nichols}
\def\secondauthor{Stefano Kalonaris}
\def\thirdauthor{Gianluca Micchi}
\def\fourthauthor{Anna Aljanaki}

\newif\ifpdf
\ifx\pdfoutput\relax
\else
   \ifcase\pdfoutput
      \pdffalse
   \else
      \pdftrue
  \fi
\fi

\ifpdf 
  \usepackage[pdftex,
    pdftitle={\papertitle},
    pdfauthor={\firstauthor, \secondauthor, \thirdauthor, \fourthauthor},
    bookmarksnumbered, 
    pdfstartview=XYZ 
   ]{hyperref}

  \usepackage[pdftex]{graphicx}
  \graphicspath{{./figures/}}
  \DeclareGraphicsExtensions{.pdf,.jpeg,.png}

  \usepackage[figure,table]{hypcap}

\else 
  \usepackage[dvips,
    bookmarksnumbered, 
    pdfstartview=XYZ 
  ]{hyperref}  

  \usepackage[dvips]{epsfig,graphicx}
  \graphicspath{{./figures/}}
  \DeclareGraphicsExtensions{.eps}

  \usepackage[figure,table]{hypcap}
\fi

\hypersetup{
    colorlinks,%
    citecolor=black,%
    filecolor=black,%
    linkcolor=black,%
    urlcolor=black
}

\title{\papertitle}

\fourauthors
   {1.5in}
   {\firstauthor} {Zillow Group, USA \\ %
     {\tt \href{mailto:epnichols@gmail.com}{epnichols@gmail.com}}}
   {\secondauthor} {Riken AIP, Japan \\ %
     {\tt \href{mailto:stefano.kalonaris@riken.jp}{stefano.kalonaris@riken.jp}}}
   {\thirdauthor} {Univ. Lille, CNRS, Centrale Lille, UMR 9189 \\CRIStAL, F-59000 Lille, France\\ %
     {\tt \href{mailto:gianluca.micchi@univ-lille.fr}{gianluca.micchi@univ-lille.fr}}}
   {\fourthauthor} {University of Tartu, Estonia \\ %
     {\tt \href{mailto:aljanaki@gmail.com}{aljanaki@gmail.com}}}

\begin{document}
\capstartfalse
\maketitle
\capstarttrue
\begin{abstract}
We propose a system for contrapuntal music generation based on a Neural Machine Translation (NMT) paradigm. 
We consider Baroque counterpoint and are interested in modeling the interaction between any two given parts as a mapping between a given source material and an appropriate target material. 
Like in translation, the former imposes some constraints on the
latter, but doesn't define it completely.
We collate and edit a bespoke dataset of Baroque pieces, use it to train an attention-based neural network model, and evaluate the generated output via BLEU score and musicological analysis. We show that our model is able to respond with some idiomatic trademarks, such as imitation and appropriate rhythmic offset, although it falls short of having learned stylistically correct contrapuntal motion ({\it e.g.}, avoidance of parallel fifths) or stricter imitative rules, such as canon.
\end{abstract}
%

\sloppy
\section{Introduction}\label{sec:introduction}
Many attempts have been made to model counterpoint ({\it e.g.}, first species \cite{doi:10.1080/17513472.2012.738554}), to formalize its rules via computational methods \cite{Song2015}, or to generate compositions in the style of four-part Bach chorales \cite{10.5555/3305381.3305522,Huang2017CounterpointBC}.
Rule and constraint-based methods have been explored extensively 
\cite{10.2307/3680335,Tsang1991HarmonizingMA}, along with grammars \cite{Gilbert2007APC,Quick:2013:GAM:2505341.2505345} and statistical methods, from Hidden Markov Models \cite{Farbood2001AnalysisAS,Allan2004HarmonisingCB} to their combinations with pattern-matching models \cite{10.2307/3680717} and templates \cite{IJIMAI2649}.
Besides these approaches, music counterpoint has also been modeled via the (increasingly more ubiquitous) machine learning paradigm \cite{Adiloglu:2007:MLA:1232940.1232984}, and specifically by the application of artificial neural networks using modern deep learning techniques, as in
\cite{10.5555/3305381.3305522,Liang2017AutomaticSC}. 

Historically, recurrent neural networks (RNNs) have often been used, since temporal dependencies are crucial in music, but more recently other techniques have also been employed, such as convolutional neural networks (CNNs), as in the \emph{Coconet} model \cite{Huang2017CounterpointBC}, and attention-based networks (see next section). Our approach also uses an attention-based neural network model. 

\section{Related Work}\label{sec:related_work}
State-of-the-art models using attention mechanisms to generate music are the \emph{Music Transformer} \cite{huang2018music}, which extends the \emph{Transformer} model \cite{10.5555/3295222.3295349} by introducing a measure of distance between any two tokens (\emph{relative attention}), and OpenAI's \emph{MuseNet} \cite{musenet}, based on the GPT-2 model \cite{gpt2}, which uses \emph{sparse attention}, whereby each of the output positions computes weightings from a subset of input positions. 

\begin{figure*}
	\centering
	\includegraphics[width=0.95\textwidth]{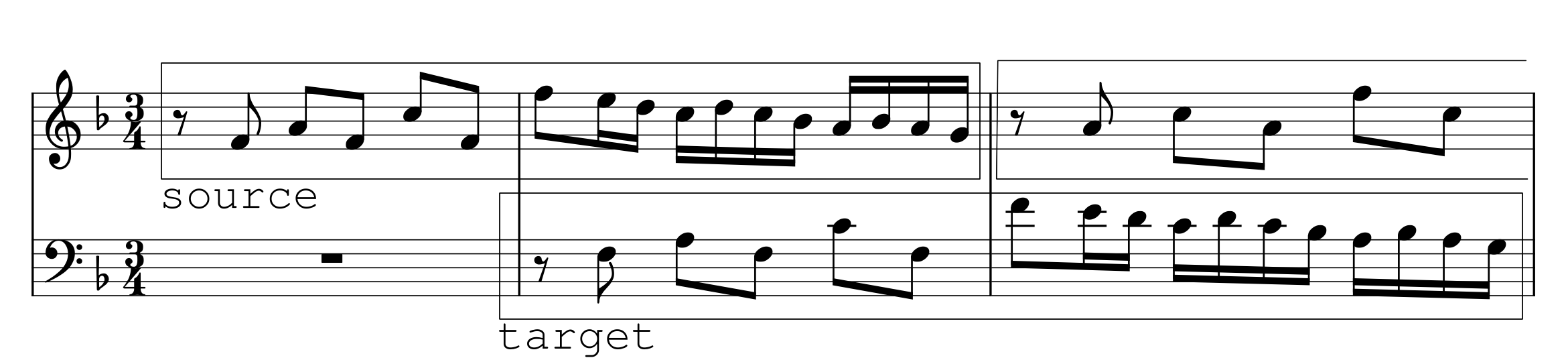}
	\caption{The incipit of J.S. Bach's two-part invention No.8, BWV 779, viewed as a NMT task.}
	\label{fig:qa}
\end{figure*}

However, both of these models encode music left-to-right and generate similarly, whereas we, instead, formulate the generation of two-part counterpoint as an NMT task (see Figure \ref{fig:qa}). In particular, we treat one of the parts as the input ({\it i.e.}, the source sentence), and train the model to generate the other part as the output ({\it i.e.}, the target sentence). 
While the use of attention-based models is certainly not new in the context of generative music systems, we contend to be the first to frame and model two-part counterpoint as a translation task, where ``translation'' means ``generating the other part''.

\section{The Data}\label{sec:data}
To train the model with suitable data, we decided to collate a bespoke dataset, motivated by the often partial and noisy nature of the readily available ones. This is an ongoing endeavor, as more pieces are continuously added, referenced against the original scores, and curated to ensure the quality of the transcription and the absence of duplicates.
Composers are chosen exclusively within the Baroque idiom ({\it e.g.}, J.S. Bach, A. Vivaldi, G.F. Handel, G.P. Telemann, S.L. Weiss, etc.) with most pieces being sourced from the Werner Icking Music Archive\footnote{\url{http://www.icking-music-archive.org/index.php}} and the Center for Computer Assisted Research in the Humanities\footnote{\url{http://kern.ccarh.org/}}.
All pieces are re-formatted as MIDI files.
At the time of writing, our dataset comprised 707 two- and three-part pieces, and 597 pieces with more than three parts, including orchestral works. 

\subsection{Data Encoding}\label{subsec:encoding}Each MIDI file is divided in tracks, one per instrument (with the exception of keyboard instruments that often have two). First of all, we removed from all individual tracks all polyphony such as doubling at the octave or occasional chords for string instruments, throwing away a single track altogether if the task was impractical or musicologically not feasible. This resulted in $k_i$ tracks per file, where $i$ indexes the files. Because we are building a model of two-part music, we built all $k_i \choose 2$ combinations of pairs of tracks. At the time of the latest version of our dataset, we had $4{,}623$ track pairs.

Next, each track pair was arbitrarily segmented into four-measure chunks and segments with less than 10 notes in any given part were filtered out in an effort to keep only the data points that are really written in a polyphonic style.
This yielded $41{,}297$ four-bar segments. Of these, $31{,}400$ were selected to make up a training set, and these segments were then transposed in all keys, augmenting the dataset to $376{,}800$ training segments and $9{,}897$ remaining segments for validation.
To encode the MIDI data for use in a neural network, we consider a vector of three elements for each note in the segment: the MIDI pitch number, the duration (floating point rounded to three decimal places), and the number of beats (also floating point, rounded) from the beginning of the segment.
We treat each piece of information as a ``word", assigning a unique string to each element. The union of these strings defines the vocabulary $V$ of the data. Each note or rest in the music is represented by a sequence of three words.
This representation affords a simple implementation of the model using existing NLP systems.

\subsection{Beat position}
    The encoding of the beat position as a language token might sound unnecessary: after all, the word embeddings to the Transformer already are composed with a global position embedding. Indeed, we could have used a beat-position embedding instead of encoding it as a token. However, we found it useful to force the model to output the correct beat position after each (variable-length) note, and noticed improved performance when the model is required to explicitly model the passage of musical time. For generating output MIDI files and for calculating BLEU scores (see Section \ref{subsec:bleu}), these beat position outputs were discarded. In a separate experiment ({\tt mod-beat-position}, below) we relied on the global position embedding from the Transformer and modified the beat position token to represent the metric position within a single measure, relative to the downbeat ({\it e.g.}, the downbeat of any measure would be encoded as position 0, the position after three eighth notes have sounded would be position 1.5, etc.).

\begin{table*}
\centering
\begin{tabularx}{0.9\textwidth}{ 
   >{\raggedright\arraybackslash}X 
   >{\centering\arraybackslash}X 
   >{\centering\arraybackslash}X 
   >{\centering\arraybackslash}X }
\toprule
\textbf{Encoding} & \textbf{Pitch} (Mean$\pm$Std) & \textbf{Duration} (Mean$\pm$Std) & \textbf{Combined} (Mean$\pm$Std)\\
\midrule
No beat-position token & 22.4$\pm$26.9 & 56.1$\pm$30.6 & 35.8$\pm$32.1 \\
{\tt beat-position} & 21.3$\pm$26.7 & \textbf{65.3}$\pm$25.3 & 38.4$\pm$33.2 \\
{\tt mod-beat-position} & \textbf{23.6}$\pm$27.4 & 63.5$\pm$25.9 & \textbf{39.6}$\pm$32.4 \\
\bottomrule
\end{tabularx}
\medskip
\caption{Our model's BLEU results.}
\label{tab:bleu}
\end{table*}

\section{The Model}\label{sec:models}
We used the OpenNMT\footnote{\url{http://opennmt.net/}} implementation of the Transformer as a basis of our model (with modifications to the beam search code).
The Transformer is made of a connected encoding and decoding network; their main components are \emph{attention} and \emph{self-attention} layers, preceded by a \emph{positional encoding} and followed by standard feed-forward layers. 
An attention layer has three inputs: a query matrix $Q$ and a pair of key-value matrices $K$ and $V$. 
In our case, for example, each row of the query matrix represents a token from the target music phrase, while each key-value pair is taken from the source music phrase.
The output of the layer is a measure of how important is each key in determining the nature of the query.
Its exact mathematical implementation can vary. The most typical one is the (modified) dot-product attention:
\begin{equation}\label{eq:attention}
\text{Attention}(Q, K, V) = \text{softmax}\left(\frac{QK^T}{\sqrt{d_k}}\right)V
\end{equation}
where $K^T$ indicates the transpose of $K$ and $d_k$ is the (common) dimension of the representation for each of the queries and keys (the other dimension being respectively the number of tokens in the query phrase and in the source phrase).
In practice, the output of the attention layer for each query is a weighted sum over all the values $V$ where the weights are given by a function measuring the mutual connection between the query and each of the keys.

In a self-attention layer, the vectors of keys, query, and values all come from the same music phrase.
If $x$ is the vector representation of the phrase (thus $x_i$ being the embedding of each token), then those three vectors are calculated as
$$ Q = x W_Q, \qquad K = x W_K, \qquad V = x W_V, $$
where $W_Q$, $W_K$, and $W_V$ are three different trainable weight matrices. Once the output of the decoder is finally calculated, a feed-forward network followed by a \emph{softmax} is used to choose the generated token out of the available ones.

The other important part of the model is the positional encoding, which determines the correct embedding of each token $x_i$ by storing all the information about the relative ordering of the tokens.
As a matter of fact, the matrix multiplications in Eq.~\ref{eq:attention} work independently on every element and disregard the ordering, treating the musical phrase ``ABCDE" equivalently to ``ADBEC".
This is solved by adding to the order-independent embedding $\tilde x_i$ a part that depends only on the position $p$ of the token in the input sequence, so that $x_i = \tilde x_i + f(p_i)$. We refer to \cite{10.5555/3295222.3295349} for details in the implementation.

The motivation of this work is that self-attention layers in the Transformer can learn musical structure by studying the relation between the different notes, for example discovering cadenzas, repetitions, and so forth.

\section{Results \& Discussion}\label{sec:results}
We evaluated our model via both NLP metrics and domain-expert opinion.

\subsection{BLEU} \label{subsec:bleu}
Extending the NMT analogy all the way from the model's architecture to the assessment of its output results, we employed the BiLingual Evaluation Understudy (BLEU) score, which is a metric used to evaluate a generated sequence against a reference sequence. BLEU is a modified precision metric over \emph{n-grams} \cite{P02-1040} (with, typically, $n \leq 4$), but it has been liable to criticism in that a sentence can be translated in many different ways. A similar argument could be made considering degree equivalence in music. For example, in the key of $C$, in a melodic phrase anchored on the pre-dominant, a $D$ (note) is contextually just as appropriate as an $F$ (note). Notwithstanding these considerations, we opted for BLEU, in the awareness that this needs to be mediated by musicological concerns. In the results below, the Pitch and Duration scores are calculated by extracting the midi pitch tokens and duration tokens, respectively, from the output stream, and then computing the BLEU score using smoothing method 2 from \cite{chencherry2014}. The Combined score was computed using the output sequence of interleaved Pitch and Duration tokens, and increasing $n$ from 4 to 8.

As a sanity check, we tested whether the model is prone to ``memorizing'' target sequences in the training data. We calculated the \emph{edit distance} \cite{Navarro:2001:GTA:375360.375365} between all possible pairs composed of a target sequence in the training set and a generated response. Edit distance was considered zero if all the pitches and their durations were identical in both sequences. We did not find any cases of direct copying behavior. For the {\tt mod-beat-position} condition, the edit distance was $19.34 \pm 9.44$ on average, while for {\tt beat-position} it was $18.43 \pm 8.25$.

The scores shown in Table \ref{tab:bleu} suggest that the {\tt mod-beat-position} version does better on the pitch-only and combined pitch+duration metrics, whereas {\tt beat-position} does better on the duration-only metric. We also computed scores for the baseline case of not using a beat position token at all; in this case, the duration-only results are much worse. 

These BLEU score results, however, do not directly translate to a measure of musical quality. We now proceed to examine, from a musicological viewpoint, some examples of model's output using both variants of the beat-position token. 

\subsection{Musical Analysis} \label{subsec:musical}
\begin{figure*}
\centering
\includegraphics[width=1\textwidth, trim = {0 1.5cm 0 1.5cm},]{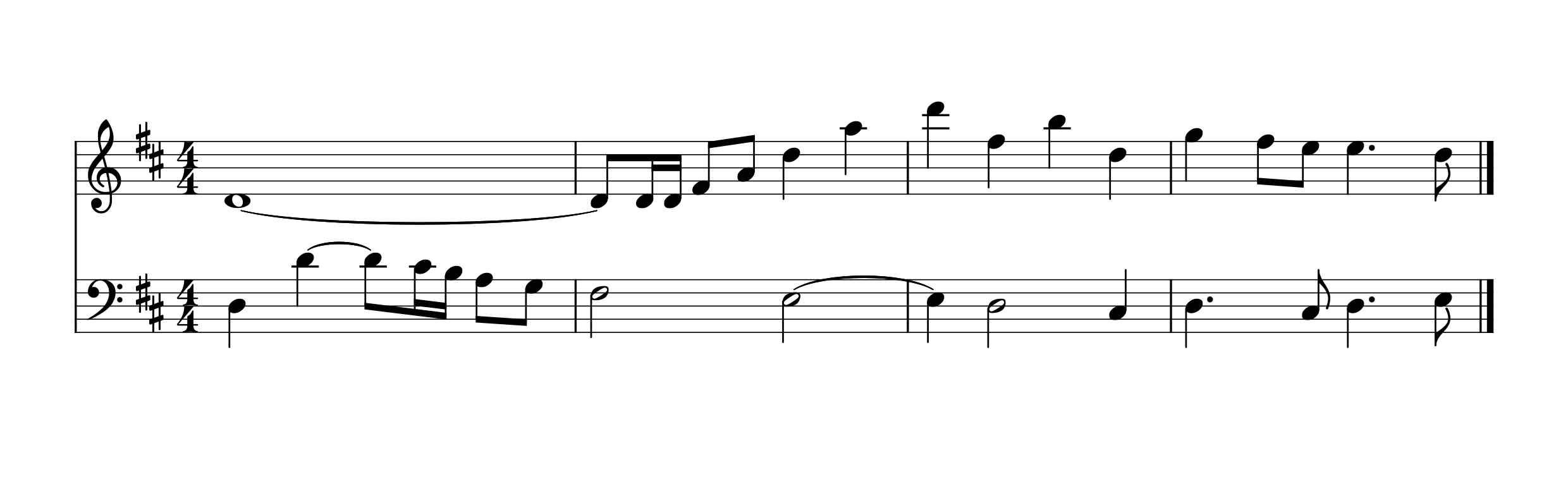}
 
\includegraphics[width=1\textwidth, trim = {0 1.5cm 0 0.5cm}]{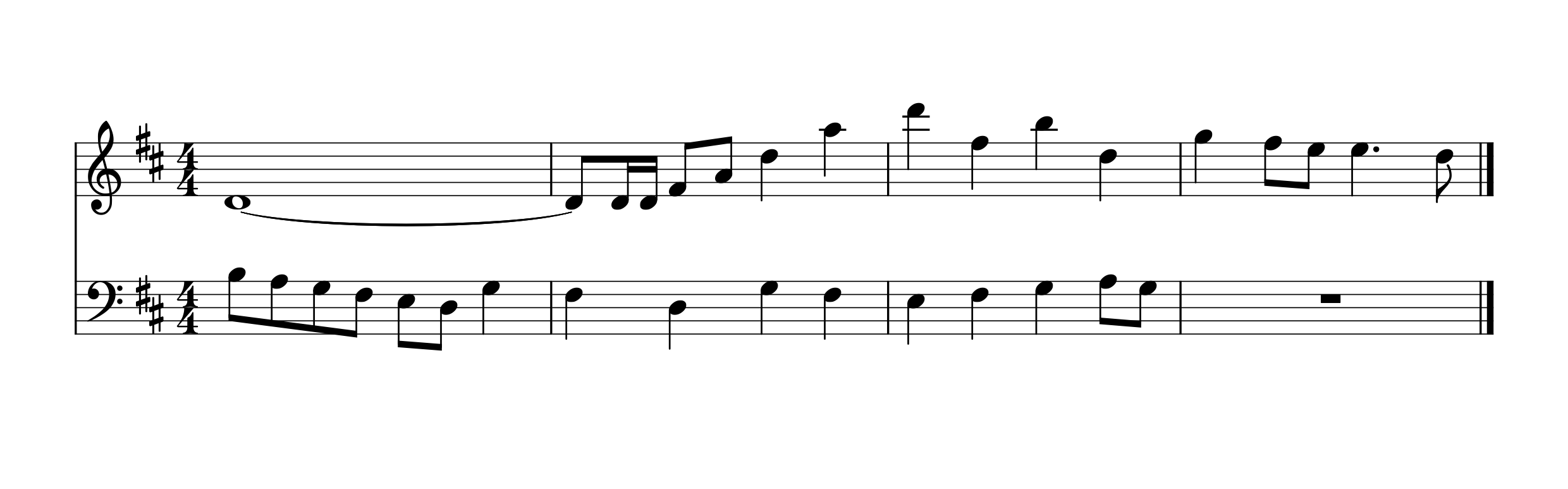}
\caption{Comparing between the {\tt mod-beat-position} (top) and the {\tt beat-position} (bottom) models' responses (bass clef) to the same query.}
\label{fig:compare}
\end{figure*}

In Figure \ref{fig:compare}, for example, we compare the model's generated responses using the {\tt mod-beat-position} and the {\tt beat-position} encoding. The former shows accomplished voice leading and it is a nice example of anticipation of the query's material, namely bars 77-80 in Handel's Messiah, movement 44 (the famous Hallelujah chorus). Moreover, the generated part imitates in contrary motion, although by diatonic steps rather than tertiary arpeggio.
When comparing this to the model's {\tt beat-position} behavior, one can notice that the rhythmic and melodic contour is more varied, comprising five (instead of two) duration values, and a wider selection of intervals, respectively. Indeed, the numerical findings previously reported seem to be corroborated by a brief musical analysis.

Despite the positive traits shown above, our model fails to exhibit certain fundamental elements of what is considered valid contrapuntal motion. Leaving aside strict applications of the  \emph{Gradus ad Parnassum} \cite{nla.cat-vn2555130} rules ({\it e.g.}, step-wise voice motion, avoiding hidden fifths and octaves, etc.), it is evident that our model produces target sequences of dubious musical appropriateness ({\it e.g.}, parallel fifths) or little harmonic coherence ({\it e.g.}, missing cadences, tonal pivots and secondary dominant leading tones). Furthermore, the model's output does not exhibit sufficient style authenticity, being, at times, more typical of the modal idiom. The most notable absence is that of canon, a more formalized type of imitation, which follows stricter rules and which comes in several guises (simple, interval, inversion, retrograde, mensuration, etc.). Sample output MIDI files are available online\footnote{\url{https://gitlab.com/skalo/baroque-nmt/-/tree/master/selected_examples}}.

\subsection{Future Work}\label{sec:todos}
An issue we foresee working on is that of hierarchical structure. While the Transformer has successfully addressed long-term structure \cite{huang2018music} (which had been the crux of many generative approaches), hierarchical modeling remains an open problem not only in NLP, where it has been shown \cite{tran-etal-2018-importance} that RNNs still outperform attention networks, but in music, too. 
We posit that hierarchical dependencies can be improved by conditioning the model on boundary segmentation of the dataset, and we intend to train the Transformer on segments obtained with perceptual \cite{10.2307/843503,Cambouropoulos01thelocal}, musicological \cite{GTTM}, and statistical \cite{marcus2005a} methods, rather than using arbitrary, fixed-length segments. 

\section{Conclusion}\label{sec:conclusion}
We presented a novel approach to Baroque counterpoint modeling, using NMT. According to this perspective, counterpoint is seen as a nearly synchronous translation task. We collated a bespoke dataset to train a Transformer model, adding a beat position token to better model musical time. We concluded that, whilst being able to generate reasonable responses at times, our model is still at odds with issues that have been long resolved in systems abiding by different architectures ({\it e.g.}, rule or constraint-based systems). Notwithstanding its current limitations, we believe that our framing of two-part polyphony is an original viewpoint worth investigating further, and we endeavor to do so in the near future.

\bibliography{bibliography}

\begin{thebibliography}{10}
\providecommand{\url}[1]{#1}
\csname url@samestyle\endcsname
\providecommand{\newblock}{\relax}
\providecommand{\bibinfo}[2]{#2}
\providecommand{\BIBentrySTDinterwordspacing}{\spaceskip=0pt\relax}
\providecommand{\BIBentryALTinterwordstretchfactor}{4}
\providecommand{\BIBentryALTinterwordspacing}{\spaceskip=\fontdimen2\font plus
\BIBentryALTinterwordstretchfactor\fontdimen3\font minus
  \fontdimen4\font\relax}
\providecommand{\BIBforeignlanguage}[2]{{%
\expandafter\ifx\csname l@#1\endcsname\relax
\typeout{** WARNING: IEEEtran.bst: No hyphenation pattern has been}%
\typeout{** loaded for the language `#1'. Using the pattern for}%
\typeout{** the default language instead.}%
\else
\language=\csname l@#1\endcsname
\fi
#2}}
\providecommand{\BIBdecl}{\relax}
\BIBdecl

\bibitem{doi:10.1080/17513472.2012.738554}
D.~Herremans and K.~S\"{o}rensen, ``Composing first species counterpoint with a
  variable neighbourhood search algorithm,'' \emph{Journal of Mathematics and
  the Arts}, vol.~6, no.~4, pp. 169--189, 2012.

\bibitem{Song2015}
X.-Y. Song and D.-R. Huang, ``{A Study on Digital Analysis of Bach's Two-Part
  Inventions},'' vol. 2015, pp. 1--6, 08 2015.

\bibitem{10.5555/3305381.3305522}
G.~Hadjeres, F.~Pachet, and F.~Nielsen, ``{DeepBach: A Steerable Model for Bach
  Chorales Generation},'' in \emph{Proceedings of the 34th International
  Conference on Machine Learning - Volume 70}, ser. ICML’17.\hskip 1em plus
  0.5em minus 0.4em\relax JMLR.org, 2017, p. 1362–1371.

\bibitem{Huang2017CounterpointBC}
C.-Z.~A. Huang, T.~Cooijmans, A.~Roberts, A.~C. Courville, and D.~Eck,
  ``{Counterpoint by Convolution},'' in \emph{Proceedings of the 18th
  International Society for Music Information Retrieval Conference}, ser.
  ISMIR'17, 2017.

\bibitem{10.2307/3680335}
K.~Ebcioğlu, ``{An Expert System for Harmonizing Four-Part Chorales},''
  \emph{Computer Music Journal}, vol.~12, no.~3, pp. 43--51, 1988.

\bibitem{Tsang1991HarmonizingMA}
C.~P. Tsang and M.~Aitken, ``{Harmonizing Music as a Discipline in Constraint
  Logic Programming},'' in \emph{Proceedings of the International Computer
  Music Conference}, ser. ICMC'91, 1991, pp. 61--64.

\bibitem{Gilbert2007APC}
{\'E}.~Gilbert and D.~Conklin, ``{A Probabilistic Context-Free Grammar for
  Melodic Reduction},'' in \emph{In International Workshop on Artificial
  Intelligence and Music, The Twentieth International Joint Conference on
  Artificial Intelligence}, ser. IJCAI'07, 2007.

\bibitem{Quick:2013:GAM:2505341.2505345}
D.~Quick and P.~Hudak, ``{Grammar-based Automated Music Composition in
  Haskell},'' in \emph{Proceedings of the First ACM SIGPLAN Workshop on
  Functional Art, Music, Modeling \& Design}, ser. FARM'13.\hskip 1em plus
  0.5em minus 0.4em\relax New York, NY, USA: ACM, 2013, pp. 59--70.

\bibitem{Farbood2001AnalysisAS}
M.~Farbood and B.~Sch{\"o}ner, ``{Analysis and Synthesis of Palestrina-Style
  Counterpoint Using Markov Chains},'' in \emph{Proceedings of International
  Computer Music Conference}, ser. ICMC'01, 2001.

\bibitem{Allan2004HarmonisingCB}
M.~Allan and C.~K.~I. Williams, ``{Harmonising Chorales by Probabilistic
  Inference},'' in \emph{Proceedings of the 17th International Conference on
  Neural Information Processing Systems}, ser. NIPS'04, 2004.

\bibitem{10.2307/3680717}
D.~Cope, ``{Computer Modeling of Musical Intelligence in EMI},'' \emph{Computer
  Music Journal}, vol.~16, no.~2, pp. 69--83, 1992.

\bibitem{IJIMAI2649}
V.~Padilla and D.~Conklin, ``{Generation of Two-Voice Imitative Counterpoint
  from Statistical Models},'' \emph{International Journal of Interactive
  Multimedia and Artificial Intelligence}, vol.~5, no.~3, pp. 22--32, 12 2018.

\bibitem{Adiloglu:2007:MLA:1232940.1232984}
K.~Adiloglu and F.~N. Alpaslan, ``{A Machine Learning Approach to Two-voice
  Counterpoint Composition},'' \emph{Knowledge-Based Systems}, vol.~20, no.~3,
  pp. 300--309, 2007.

\bibitem{Liang2017AutomaticSC}
F.~T. Liang, M.~Gotham, M.~Johnson, and J.~Shotton, ``{Automatic Stylistic
  Composition of Bach Chorales with Deep LSTM},'' in \emph{Proceedings of the
  18th International Society for Music Information Retrieval Conference}, ser.
  ISMIR'17, 2017.

\bibitem{huang2018music}
C.-Z.~A. Huang, A.~Vaswani, J.~Uszkoreit, N.~Shazeer, C.~Hawthorne, A.~M. Dai,
  M.~D. Hoffman, and D.~Eck, ``{Music Transformer: Generating Music with
  Long-Term Structure},'' \emph{arXiv preprint arXiv:1809.04281}, 2018.

\bibitem{10.5555/3295222.3295349}
A.~Vaswani, N.~Shazeer, N.~Parmar, J.~Uszkoreit, L.~Jones, A.~N. Gomez,
  L.~Kaiser, and I.~Polosukhin, ``Attention is All You Need,'' in
  \emph{Proceedings of the 31st International Conference on Neural Information
  Processing Systems}, ser. NIPS’17.\hskip 1em plus 0.5em minus 0.4em\relax
  Red Hook, NY, USA: Curran Associates Inc., 2017, p. 6000–6010.

\bibitem{musenet}
``{MuseNet},'' \url{https://openai.com/blog/musenet/}, {A}ccessed: 2019-06-27.

\bibitem{gpt2}
\BIBentryALTinterwordspacing
A.~Radford, J.~Wu, R.~Child, D.~Luan, D.~Amodei, and I.~Sutskever, ``{Language
  Models are Unsupervised Multitask Learners},'' 2018. [Online]. Available:
  \url{https://d4mucfpksywv.cloudfront.net/better-language-models/language-models.pdf}
\BIBentrySTDinterwordspacing

\bibitem{P02-1040}
K.~Papineni, S.~Roukos, T.~Ward, and W.-J. Zhu, ``{Bleu: a Method for Automatic
  Evaluation of Machine Translation},'' in \emph{Proceedings of 40th Annual
  Meeting of the Association for Computational Linguistics}.\hskip 1em plus
  0.5em minus 0.4em\relax Philadelphia, Pennsylvania, USA: Association for
  Computational Linguistics, Jul. 2002, pp. 311--318.

\bibitem{chencherry2014}
B.~Chen and C.~Cherry, ``A Systematic Comparison of Smoothing Techniques for
  Sentence-Level {BLEU},'' in \emph{Proceedings of the Ninth Workshop on
  Statistical Machine Translation}.\hskip 1em plus 0.5em minus 0.4em\relax
  Baltimore, Maryland, USA: Association for Computational Linguistics, Jun.
  2014, pp. 362--367.

\bibitem{Navarro:2001:GTA:375360.375365}
G.~Navarro, ``A Guided Tour to Approximate String Matching,'' \emph{ACM
  Computing Surveys}, vol.~33, no.~1, pp. 31--88, Mar. 2001.

\bibitem{nla.cat-vn2555130}
J.~J. Fux, A.~Mann, and J.~Edmunds, \emph{{The study of counterpoint from
  Johann Joseph Fux's Gradus ad parnassum. Translated and edited by Alfred
  Mann, with the collaboration of John Edmunds}}, rev.~ed.\hskip 1em plus 0.5em
  minus 0.4em\relax W. W. Norton New York, 1965.

\bibitem{tran-etal-2018-importance}
K.~Tran, A.~Bisazza, and C.~Monz, ``The Importance of Being Recurrent for
  Modeling Hierarchical Structure,'' in \emph{Proceedings of the 2018
  Conference on Empirical Methods in Natural Language Processing}.\hskip 1em
  plus 0.5em minus 0.4em\relax Brussels, Belgium: Association for Computational
  Linguistics, Oct.-Nov. 2018, pp. 4731--4736.

\bibitem{10.2307/843503}
J.~Tenney and L.~Polansky, ``{Temporal Gestalt Perception in Music},''
  \emph{Journal of Music Theory}, vol.~24, no.~2, pp. 205--241, 1980.

\bibitem{Cambouropoulos01thelocal}
E.~Cambouropoulos, ``{The Local Boundary Detection Model (LBDM) and its
  application in the study of expressive timing},'' in \emph{Proceedings of the
  International Computer Music Conference}, ser. ICMC'01, 2001.

\bibitem{GTTM}
F.~Lerdahl and R.~Jackendoff, \emph{{A generative theory of tonal
  music}}.\hskip 1em plus 0.5em minus 0.4em\relax Cambridge, MA: MIT Press,
  1983.

\bibitem{marcus2005a}
M.~Pearce, ``{The Construction and Evaluation of Statistical Models of Melodic
  Structure in Music Perception and Composition},'' Ph.D. dissertation, School
  of Informatics, City University, London, 2005.

\end{thebibliography}

\end{document}